\documentstyle[twoside,fleqn,espcrc2]{article}

\newcommand{\bee} {\begin{equation}}
\newcommand{\ene} {\end{equation}}
\newcommand{\bea} {\begin{array}}
\newcommand{\ena} {\end{array}}
\newcommand{\beqa} {\begin{eqnarray}}
\newcommand{\enqa} {\end{eqnarray}}
\newcommand{\lapproxeq}{\lower .7ex\hbox{$\;\stackrel{\textstyle <}{\sim}\;$}}
\def\al{\alpha}
\def\si{\sin^2{\theta_W(M_Z)}}
\def\tiz{SU(4)_{PS}\otimes SU(2)_L\otimes SU(2)_R}
\def\lu{SU(3)_c\otimes SU(2)_L\otimes SU(2)_R\otimes U(1)_{B-L}}
\def\sm{SU(3)_c\otimes SU(2)_L\otimes U(1)_Y}
\def\SM{G_{SM}}
\def\ul{SU(3)_c\otimes U(1)_Q}
\def\rw{\rightarrow}
\def\lrw{\longrightarrow}
\def\t{\tau_{p\rw e^+\pi^0}}

\newcommand{\AmS}{{\protect\the\textfont2
  A\kern-.1667em\lower.5ex\hbox{M}\kern-.125emS}}

\title{SO(10): A possible scenario for new physics in the neutrino sector
       and baryogenesis}

\author{
  G. Amelino-Camelia\address{Center for Theoretical Physics, Laboratory
    of Nuclear Science and Department of Physics, \\ Massachusetts Institute
    of Technology, Cambridge, Massachusetts 02139, USA.},
  O. Pisanti\address{Dipartimento di Scienze Fisiche, Mostra
    d'Oltremare Pad. 19, 80125 Napoli, Italia. \\ Istituto Nazionale di
    Fisica Nucleare, Mostra d'Oltremare Pad. 20, 80125 Napoli, Italia.},
  and L. Rosa$^b$
       }

\begin{document}

\begin{abstract}
The implications on neutrino physics and on the dynamical generation of the
baryonic asymmetry of a class of $SO(10)$ non-supersymmetric models are
discussed.
\end{abstract}

\maketitle

\section{INTRODUCTION}

In~the~last~years~$SO(10)$~non-supersymmetric GUT models \cite{frmi} have
been the subject of renewed attention because people recognized
they may represent a Standard Model (SM) extension in which the unification
of the strong, electromagnetic and weak interactions is consistent with the
experimental values of $\al(M_Z)$, $\si$ and $\al_S(M_Z)$.
Moreover, they can give, through the see-saw mechanism \cite{gera}, neutrino
masses of the order required to explain the solar-neutrino problem within the
framework of the MSW theory \cite{msw}, and to account for the baryon
asymmetry \cite{mang} and for at least part of the dark matter of the universe
\cite{schr}.

In this paper we will describe the results of a research on four particular
$SO(10)$ symmetry breaking patterns. By studying the Higgs potential of the
model, and using the Renormalization Group Equations (RGE), it was possible
to deduce \cite{lone} the values of the two physical
scales of the theory: the unification scale, $M_X$, at which $SO(10)$
breaks to an intermediate group $G'$ (in the cases under investigation
always greater than the standard group, {\small $\sm\equiv\SM$}),
and the intermediate scale, $M_R$, at which the intermediate group is
broken to the standard group, {\small $\SM$}. Therefore, a typical $SO(10)$
breaking chain is given by

{\small
\bee
SO(10)\buildrel M_X\over\lrw G'\buildrel M_R\over\lrw\SM\buildrel M_Z\over
\lrw\ul. \label{eq:ssb}
\ene
}
\noindent $M_X$ is connected with the masses of the lepto-quarks which
mediate proton decay, and, in particular, the present experimental lower limit
on proton decay, $\t\geq 9\cdot10^{32}~years$ \cite{padb}, corresponds
\cite{oliv} to the following lower limit on $M_X$:
\bee
M_X\geq 3.2\cdot10^{15}~GeV. \label{eq:mxinf}
\ene
Through the see-saw mechanism, $M_R$ is related to the masses of the (almost)
left-handed neutrinos.

\section{$SO(10)$ GUT MODELS}

$SO(10)$ unified models have been studied since many years with the
physical motivation of obtaining values for the masses of the
lepto-quarks which mediate proton decay higher than the ones found within
the $SU(5)$ minimal model.

Recently, the more precise determination of the gauge coupling constants at
the scale $M_Z$ has allowed to show that, if only standard model particles
contribute to the RGE, the three running coupling constants of {\small $\SM$}
meet at three different points \cite{amal}
and only the meeting
point of $\al_2(\mu)$ and $\al_S(\mu)$ corresponds to a value of the scale
$\mu$ sufficiently high to comply with the experimental lower limit on proton
decay.

$SO(10)$, in which the hypercharge is the combination
of two generators belonging to its Cartan,
\bee
Y=T_{3R}+\frac{B-L}{2},
\ene
where $T_{3R}$ and $B-L$ belong to $SU(2)_R$ and $SU(4)_{PS}$ respectively,
is very promising to modify the SU(5) predictions in such a way to prevent
conflict with experiment. In fact, if there is an intermediate symmetry
group $G'$ containing $SU(2)_R$ and/or $SU(4)_{PS}$, it is possible to
substitute the Abelian evolution of $Y$ with the non-Abelian one of either
component of $Y$, getting a higher unification point.

$SO(10)$ also reduces more than $SU(5)$ the amount of arbitrariness which
characterizes the SM. First of all, it accommodates in one irreducible
representation (IR), the spinorial one of dimension 16, the 15 known
left-handed fermions of a generation plus a new particle, whose
quantum numbers are the same as those of the not yet discovered $\nu_L^c$ .
In this way, a real unification is realized with respect to the reducible
representation $\bar{5}+10$ which accommodates the fermions in the minimal
$SU(5)$.
Moreover, the presence of $\nu_L^c$ leads to a mass matrix, for one
generation of neutrinos, of the form
\beqa
  \bea{lll}
    \bea{lr}
      \left(\overline{\nu_R^c} \right. & \left.\overline{\nu_R}\right) \\
                                       &
    \ena\!\!\!\!\!
    &
    \!\!\!\!\left(
    \bea{lr}
      0 & \frac{m_D}{2} \\
      \frac{m_D}{2} & M
    \ena
    \right)\!\!\!
    &
    \!\!\!\left(
    \bea{c}
    \nu_L \\
    \nu_L^c
    \ena
    \right)
  \ena\!\!\! +~h.c.,
\label{eq:masmat}
\enqa
where $m_D$ is the Dirac mass and $M$ the Majorana mass of the right-handed
neutrinos. If $m_D\ll M$, the hypothesis on which the see-saw mechanism
relies, the diagonalization of the mass matrix in Eq.~(\ref{eq:masmat})
gives the two eigenvalues
\bee
m_{\nu_1}\sim \frac{m_D^2}{M},~~m_{\nu_2}\sim M,
\ene
and the relation $m_{\nu_1}\ll m_D$ agrees with the observation that
the neutrinos have a mass, if any, much smaller than the one of the other
fermions.

{\footnotesize
\begin{table}[tbh]
\setlength{\tabcolsep}{0.05pc}
\newlength{\ddigitwidth} \settowidth{\ddigitwidth}{\rm 0}
\caption{Classification of the Higgs $\SM$-invariant components in
the lower irreducible representation of $SO(10)$. $\SM$ is the standard
group.}
\label{tab:class}
\begin{tabular}{@{}rcl}
\hline
                 \multicolumn{1}{c}{IR}             &
                 \multicolumn{1}{c}{$\SM$}          &
                 \multicolumn{1}{c}{Symmetry}          \\
           &     \multicolumn{1}{c}{-singlet}       &  \\
\hline
16   &  $\hat{\chi}$    &  $SU(5)$                                          \\
45   &  $\hat{\al_1}$   &  $\lu\times D$                                    \\
45   &  $\hat{\al_2}$   &  $SU(4)_{PS}\otimes SU(2)_L\otimes U(1)_{T_{3R}}$ \\
54   &  $\hat{\sigma}$  &  $\tiz\times D$                                   \\
126  &  $\hat{\psi}$    &  $SU(5)$                                          \\
144  &  $\hat{\omega}$  &  $\sm$                                            \\
210  &  $\hat{\phi_1}$  &  $\lu\times D$                                    \\
210  &  $\hat{\phi_2}$  &  $\tiz$                                           \\
210  &  $\hat{\phi_3}$  &  $SU(3)_c\otimes SU(2)_L\otimes U(1)_{T_{3R}}\otimes
U(1)_{B-L}$                                                                 \\
\hline
\end{tabular}
\end{table}
}

Another well known feature implied by the choice of $SO(10)$ as a gauge
group is the absence of triangle anomalies, due to the fact that in $SO(10)$
it is not possible to construct a cubic invariant with the adjoint
representation which the gauge bosons belong to (In $SU(5)$ this results
from an accidental cancellation of the $\bar{5}$ and $10$ contributions.).

\section{THE SPONTANEOUS SYMMETRY BREAKING OF $SO(10)$}

In order to identify the possible directions for the Spontaneous Symmetry
Breaking (SSB) of $SO(10)$, one has to classify the components, in the
smallest IR's of the group, that are invariant under $\SM$.
{}From the classification in Table~\ref{tab:class}, where D stands for the
left-right discrete symmetry which interchanges $SU(2)_L$ and $SU(2)_R$
\cite{kuzm}, it is possible to understand the reason why the $SO(10)$
breaking chain typically has one more step than the $SU(5)$ one: indeed, we
see that for all the IR's, but the 144, the little group of the
$\SM$-singlet is greater than $\SM$.

Actually, either for phenomenological or for technical reasons, some of the
directions in Table~\ref{tab:class} cannot be used for the first spontaneous
breaking step. The use of the 16, 126 and 144 representations would lead to
the result that, like in $SU(5)$ GUT's, the three SM running coupling constants
do not meet at the same point. Concerning the 45 representation, one can
show that the only non-trivial positive definite invariant with degree $\leq 4$
(as necessary in order to have a renormalizable potential) that one
can build has its minimum in the {\small $SU(5) \otimes U(1)$}-invariant
and its maximum in the {\small $SO(8) \otimes SO(2)$}-invariant directions
\cite{ext45}, so that it is not possible to construct a Higgs potential
with minimum along the $\hat{\al_1}$ or $\hat{\al_2}$ directions.
Moreover, the $\hat{\phi_3}$ component in the 210 representation
corresponds to a direction with neither {\small $SU(4)_{PS}$} nor {\small
$SU(2)_R$} in the little group.

The previous considerations lead us to the following four patterns, in
which the first steps are:

{\footnotesize
\[
  \bea{ll}
    SO(10) &                  \\ \\
    ~~(i)\buildrel\hat{\sigma}\over\lrw   & \tiz\times D  \\ \\
    ~(ii)\buildrel\hat{\phi_1}\over\lrw   & \lu\times D   \\ \\
    (iii)\buildrel\hat{\phi_2}\over\lrw   & \tiz          \\ \\
    (iv)\buildrel\hat{\phi_4}\over\lrw   & \lu,
  \ena
\]
}
\noindent where $\hat{\phi_4}= cos\theta\hat{\phi_1} + sin \theta\hat{\phi_2}$.

The other steps of the SSB (see Eq.~(\ref{eq:ssb})) are common to the four
patterns. The second one is realized using the $\hat{\psi}$-component of
a $126\oplus\overline{126}$ representation, and the third one by a combination
of the {\small $\ul$}-invariant components of two 10's, in such a way to
avoid the unwanted relation $m_t=m_b$ \cite{lone}.

The previous four possibilities have been studied in the
Refs.\cite{lore,lui,tiz,luid}\footnote{Specifically, (i) has been
analyzed in Ref.\cite{lore}, (ii) in Ref.\cite{lui}, (iii) in Ref.\cite{tiz},
and (iv) in Ref.\cite{luid}.}. In most of these
papers the models have been investigated within the Extended Survival
Hypothesis (ESH), which assumes that the Higgs scalars acquire their masses
at the highest possible scale whenever this is not forbidden by symmetries
\cite{esh}. However, as discussed in Ref.\cite{lone}, the ESH may be too
drastic since in the 210 and 126
representations of $SO(10)$ there are multiplets with high quantum numbers,
which may give important contributions to the RGE. On the opposite side, if
one assumes total freedom \cite{dish} in assigning masses to the Higgs
scalars, huge uncertainties are introduced in the $SO(10)$-predictions.
However, this assumption of total freedom is also too drastic; in fact,
the mass spectrum of the Higgs scalars depends on the
coefficients of the non-trivial invariants that appear in the scalar
potential, which are constrained by the condition that the absolute minimum of
the potential is in the direction giving the desired symmetry breaking pattern
\cite{lone}.
In Ref.\cite{lone}, we released the ESH, but, by taking into account the
just mentioned constraints on the Higgs spectra, we were able to derive some
restrictive conditions on the contributions of the Higgs scalars to the RGE,
and showed that the resulting uncertainties on the $SO(10)$-predictions are
much smaller than the ones expected by Ref.\cite{dish}. Then, for each of
the four models under investigation, we searched for the upper limit on the
intermediate scale, $M_R^{UP}$, corresponding to a lower limit on the neutrino
masses, and evaluated the corresponding value of the unification scale, $M_X$
\cite{lone}.

We report here only the results for the case {\small $G'=\lu$}, which is
the most interesting one:
\bee
\bea{l}
M_R^{UP}=1.2\cdot 2.5^{0\pm 1}\cdot 10^{11}~GeV, \\ \\
M_X=1.9\cdot 2.0^{0\pm 1}\cdot 10^{16}~GeV.
\ena
\ene

Within the see-saw mechanism, the upper limit for $M_R$ gives rise to the
following inequalities for $m_{\nu_\tau}$ and $m_{\nu_\mu}$:
\bee
\bea{lll}
m_{\nu_{\tau}} &\geq& 11~ \frac{g_{2R}(M_R)}{f_3(M_R)}
\left(\frac{m_t}{100~GeV}\right)^2~eV  \\ \\
m_{\nu_{\mu}}  &\geq& 2.4\cdot10^{-3}~\frac{g_{2R}(M_R)}{f_2(M_R)}~eV,
\ena
\label{eq:mnu}
\ene
where $g_{2R}$ and $f_i$ are the $SU(2)_R$ gauge coupling constant and the
Yukawa coupling of the $126\oplus\overline{126}$ to the i-th family
respectively.
For natural values of $g_{2R}$ and $f_i$, Eqs.~(\ref{eq:mnu})
imply a substantial contribution of $\nu_\tau$ to the dark
matter in the universe and a $m_{\nu_\mu}$ relevant for the MSW solution
of the solar-neutrino problem.

\section{$SO(10)$ BARYOGENESIS}

Another interesting possible prediction of this class of $SO(10)$ models is a
dynamical explanation of the presently observed baryon asymmetry. Indeed,
these models can satisfy the three necessary conditions stated by Sakharov
\cite{sakh}: i) the $SO(10)$-gauge bosons mediate interactions which may lead
to B-violations; ii) at the intermediate scale, C and CP symmetry are
broken\footnote{No $\Delta B$ can be generated
until C and CP symmetry remain unbroken, and this happens only at $M_R$ since
the intermediate group has the same rank of $SO(10)$ \cite{mang}.}; iii)
non-equilibrium conditions can be implemented if the masses of the Higgs
particles satisfy certain conditions.

A scenario in which i), ii), and iii) are realized is discussed in
Ref.\cite{mang}, in which our $SO(10)$ model with {\small $G'=\lu$} is
considered. A non-zero value of $\Delta (B-L)$ is produced at $T\lapproxeq
M_R$ by the $B-L$-violating decays $\tilde{\phi}\rw\tilde{\psi} f \nu_L^c$,
where $\tilde{\phi}$ are some Higgs multiplets of the 210 described in
Ref.\cite{mang}, $\tilde{\psi}$ are the Higgs of the 126 which have mass of
order $M_R$, and $f$ is a fermion. The knowledge of the mass spectrum of
the Higgs scalars allowed the authors of Ref.\cite{mang} to verify the
possibility to have an overabundant population of $\tilde{\phi}$ at $M_R$,
expressed by the inequalities $10^{12}GeV\leq m_{\tilde{\phi}}\leq 4\cdot
10^{14}GeV$, where the first and second inequality correspond to the condition
that the annihilation and decay processes respectively are "frozen out".
If no $B-L$-violating phenomena is active at
lower temperatures, the stored $\Delta (B-L)$ is transformed in $\Delta B$ by
the sphaleronic processes\footnote{Note that the problem of the first
proposals of $SO(10)$-based baryogenesis was that the sphaleronic processes
would wash out the stored $\Delta B$, but in the $SO(10)$-based baryogenesis
here considered the sphaleronic processes play the positive role of
transforming the stored $\Delta (B-L)$ in $\Delta B$.}.

\section{CONCLUSIONS}

We find that $SO(10)$ phenomenology can quite naturally
accomodate non-conventional neutrino physics and a dynamical mechanism for
the generation of the baryonic asymmetry, and therefore these models may
play an important role in future developments of astroparticle physics.

\section*{ACKNOWLEDGEMENTS}

One of us, O. P., would like to thank Dr. G. Mangano for useful discussions.


\begin{thebibliography}{9}
\bibitem{frmi} H. Georgi, Particles and Fields, C. E. Carlson, AJP, 1975;
H. Fritzsch and P. Minkowski, Ann. of Phys. 93 (1975) 183.
\bibitem{gera} M. Gell-Mann, P. Ramond, and R. Slansky, Supergravity,
North Holland, Amsterdam, 1980; T. Yanagida, Proceedings of the Workshop
on the Unified Theory and the Baryon Number of the Universe, edited by O.
Sawada et al., KEK, 1979.
\bibitem{msw} L. Wolfenstein, Phys. Rev. D 17 (1978) 2369; S. P. Mikheyev
and A. Y. Smirnov, Yad. Fiz. 42 (1985) 1441 [Sov. J. Nucl. Phys. 42 (1985)
913].
\bibitem{mang} F. Buccella, G. Mangano, A. Masiero, and L. Rosa, Phys.
Lett. B 320 (1994) 313.
\bibitem{schr} D. Schramm, Nucl. Phys. B (Proc. Suppl.) 28 A (1992) 243.
\bibitem{lone} F. Acampora, G. Amelino-Camelia, F. Buccella, O. Pisanti,
L. Rosa, and T. Tuzi, to be published in Il Nuovo Cimento, Sect. A.
\bibitem{padb} Review of Particle Properties, Phys. Rev. D 45 n. 11 (1992).
\bibitem{oliv} A. le Yaouanc, L. Oliver, O. P\`ene, and J. C. Raynal,
Phys. Lett. B 72 (1977) 53.
\bibitem{amal} U. Amaldi, W. de Boer, and H. Furstenau, Phys. Lett. B 260
(1991) 447.
\bibitem{kuzm} V. Kuzmin and N. Shaposhnikov, Phys. Lett. B 92 (1980) 115.
\bibitem{ext45} L. Li, Phys. Rev. D 9 (1974) 1723.
\bibitem{lore} Q. Shafi and C. Wetterich, Phys. Lett. B 85 (1979) 52;
F. Buccella, L. Cocco, and C. Wetterich, Nucl. Phys. B 243 (1984) 273;
M. Abud, F. Buccella , A. Della Selva, A. Sciarrino, R. Fiore, and G.
Immirzi, Nucl. Phys. B 263 (1986) 336.
\bibitem{lui} F. Buccella and L. Rosa, Zeitschrift f$\ddot u$r Phys. C 36
(1987) 425.
\bibitem{tiz} D. Chang, R. N. Mohapatra, and M. K. Parida, Phys. Rev. Lett.
52 (1984) 1072; D. Chang, J. M. Gipson, R. E. Marshak, R. N. Mohapatra, and
M. K. Parida, Phys. Rev. D 31 (1985) 1718; F. Buccella, L. Cocco, A. Sciarrino,
and T. Tuzi, Nucl. Phys. B 274 (1986) 559; R. N. Mohapatra and M. K. Parida,
Phys. Rev. D 47 (1993) 264.
\bibitem{luid} M. Abud, F. Buccella, L. Rosa, A. Sciarrino, Zeitschrift
$f\ddot u$r Phys. C 44 (1989) 589; F. Buccella and L. Rosa, Nucl. Phys. B
(Proc. Suppl.) 28 A (1992) 168.
\bibitem{esh} F. del Aguila and L. E. Ibanez, Nucl. Phys. B 177 (1981) 60.
\bibitem{dish} V. V. Dixit and M. Sher, Phys. Rev. D 40 n.11 (1989) 3765.
\bibitem{sakh} A. D. Sakharov, Pis'ma Zh. Eksp. Teor. Fiz. 5 (1967) 32
[JETP Lett. 5 (1967) 24].
\end{thebibliography}
\end{document}